\documentclass[prl
,twocolumn%
,aps]{revtex4}
\usepackage{natbib}
\usepackage{graphicx}
\begin{document}
\title{Counter-Intuitive Vacuum-Stimulated Raman Scattering}
\author{Markus Hennrich, Thomas Legero, Axel Kuhn, and Gerhard Rempe}
\affiliation{Max-Planck-Institut f\"{u}r Quantenoptik, Hans-Kopfermann-Str.1, 85748 Garching, Germany}
\date{\today}
\begin{abstract}
Vacuum-stimulated Raman scattering in strongly coupled atom-cavity systems allows one to generate free-running single photon pulses on demand. Most properties of the emitted photons are well defined, provided spontaneous emission processes do not contribute. Therefore, electronic excitation of the atom must not occur, which is assured for a system adiabatically following a dark state during  the photon-generation process. We experimentally investigate the conditions that must be met for adiabatic following in a time-of-flight driven system, with atoms passing through a cavity and a pump beam oriented transverse to the cavity axis. From our results, we infer the optimal intensity and relative pump-beam position with respect to the cavity axis.
\end{abstract}
\maketitle
During the last years, many different types of single-photon sources have been proposed and successfully demonstrated \cite{Law96,Law97,Kim99,Lounis00,Kurtsiefer00,Brouri00,Michler00,Santori01,Yuan02,Michler00:2,Moreau01,Kuhn99,Hennrich00,Kuhn01,Kuhn02}. These activities were motivated by current quantum-cryptography schemes \cite{Tittel98}, which rely on the transmission of single photons between two parties to be secure, and by recent proposals \cite{Monroe02} on all-optical quantum information processing \cite{Knill01} and distributed quantum networking \cite{Cirac97}.
For quantum networking, indistinguishable photons must be generated by a unitary, reversible process in order to transport single quantum bits from node to node. This last requirement is not met by most of the single-photon sources available so far.
Only photon-generation schemes that are based on a deterministic and unitary energy exchange between a single emitter, e.g. an atom, and a single mode of the quantized radiation field \cite{Cirac97,Enk97:1,Parkins93,Parkins95,Pellizari95}
might, in principle, meet the above requirements. Such a deterministic energy exchange has been successfully demonstrated in the microwave regime \cite{Rauschenbeutel99,Brattke01}, but these experiments do not provide free-running photons.

\begin{figure}
\includegraphics[width=8.5cm]{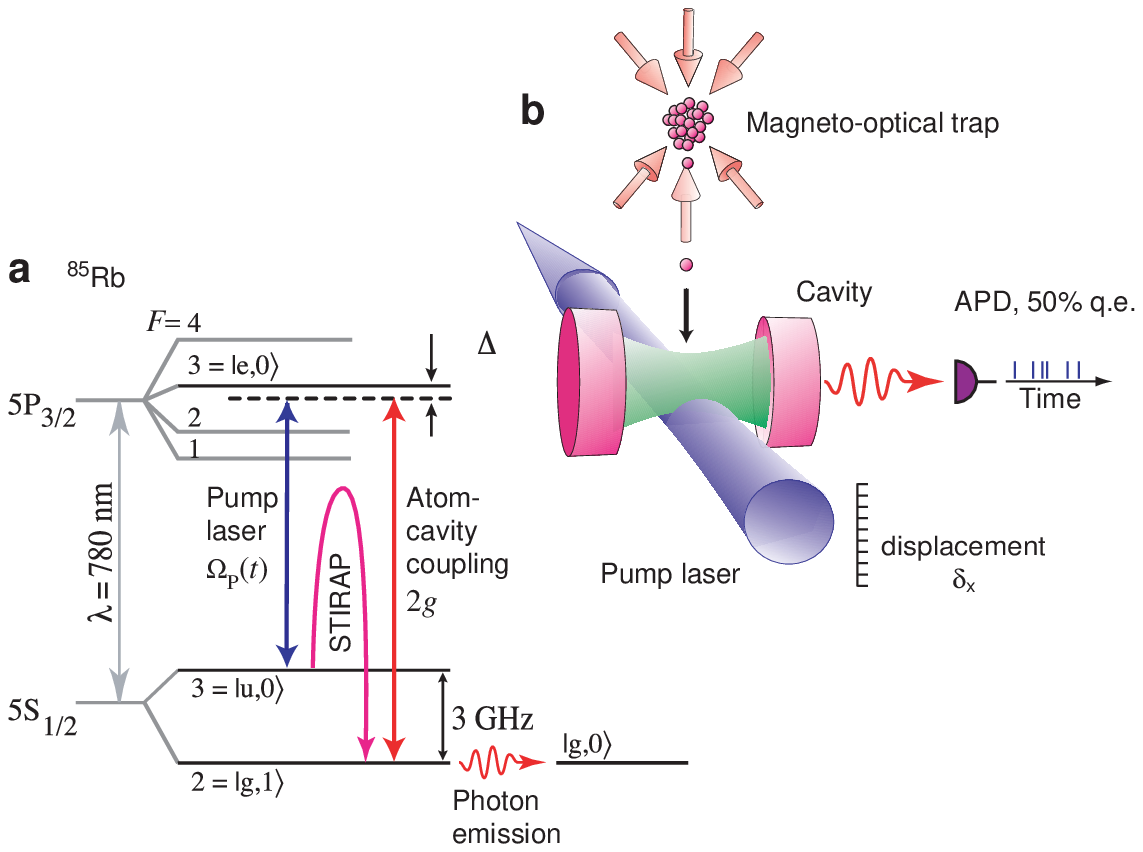}
\caption{\label{kfig1}Experimental scheme. \textbf{(A)} Energy 
levels and transitions in $^{85}$Rb. The states labeled 
$\left| u\right\rangle$, 
$\left| e\right\rangle $
 and 
$\left| g\right\rangle $
 are involved in the process, and states 
$\left| 0\right\rangle $
 and 
$\left| 1\right\rangle $
 denote the photon number in the cavity. \textbf{(B)} Setup: A cloud 
of atoms is released from a magneto-optical trap and falls through 
a cavity $20\,$cm below in about $8\,$ms with a velocity 
of $2\,$m/s. The interaction time of a single atom with the 
TEM$_{00}$ mode of the cavity (waist $w_{C}=35\,\mu$m) 
amounts to about $17.5\,\mu$s. The pump laser partially overlaps with the cavity mode. Photons emitted from the cavity are detected by an avalanche photodiode 
with a quantum efficiency of 50\%.}
\end{figure}

In this paper, we concentrate on a strongly coupled atom-cavity system in the optical domain, where the photons are transmitted through one of the cavity mirrors.  The photons are generated by a vacuum-stimulated Raman transition \cite{Kuhn99,Hennrich00,Kuhn01,Kuhn02} that is adiabatically driven by a pump laser beam, interacting with the atoms while they are coupled to the cavity mode. The excitation scheme implements stimulated Raman scattering by adiabatic passage \cite{Vitanov01} (STIRAP), with the vacuum field of the cavity acting instead of a stimulating laser. Figure\,\ref{kfig1} illustrates this scheme: A cloud of Rubidium atoms in state $|u\rangle$, i.e. the $5S_{1/2}(F=3)$ hyperfine state of the electronic ground state, is released from a magneto-optical trap (MOT). The atoms fall through a 1\,mm long cavity of Finesse $F=60\,000$ which is placed 20\,cm below the  MOT, so that the atoms passing through the cavity have acquired a transverse (vertical) velocity of $v=2\,$m/s. One TEM$_{00}$ mode of the cavity with waist $w_{C}=35\,\mu$m stimulates the Raman process, since it is resonant with the $|e\rangle\leftrightarrow|g\rangle$ transition of the atom, with $|e\rangle$ being an electronically excited state, and $|g\rangle$ being the other hyperfine ground state, $5S_{1/2}(F=2)$. The transition is pumped by a continuous laser beam of waist $w_{P}\approx w_{C}$, resonant with the $|u\rangle\leftrightarrow|e\rangle$ transition, that perpendicularly crosses the cavity axis and the trajectory of the falling atoms. With $\delta_{X}$ being the distance between the crossing points of cavity and beam axis, respectively, with the atomic trajectory, the Rabi frequency of the pump beam driving the $|u\rangle\leftrightarrow|e\rangle$ transition, $\Omega_{P}$, and the vacuum-Rabi frequency induced by the cavity coupling the $|e\rangle\leftrightarrow|g\rangle$ transition, $\Omega_{C}$, become time dependent and delayed due to the motion of the atoms, according to
\begin{eqnarray}
\Omega_{P}(t)&=&\Omega_{0} e^{-\left(\frac{vt+\delta_{X}}{w_{P}}\right)^2} 
\quad\mbox{and}\\
\Omega_{C}(t)&=&2g(t)=2g_{0} e^{-\left(\frac{vt}{w_{C}}\right)^2},\nonumber
\label{pshape}
\end{eqnarray}
where $\Omega_{0}$ is the peak Rabi frequency of the pump beam and $g_{0}$ the peak atom-cavity coupling constant.
In case of positive $\delta_{X}$, the atoms are first exposed to the pump beam, while the interaction with the cavity mode is delayed by $\delta_{X}/v$. The opposite situation is met in case of negative $\delta_{X}$. If we consider a Raman-resonant excitation with the detunings, $\Delta$, of the cavity and the pump beam from their respective transitions being equal, the interaction Hamiltonian of the coupled laser-atom-cavity system reads (in a rotating frame)
\begin{equation}
H_\mathrm{int}= \frac{\hbar}{2} \left[2\Delta \sigma_{ee} - 2g (\sigma_{eg} a + a^\dag \sigma_{ge}) 
- \Omega_{P} (\sigma_{eu} + \sigma_{ue})\right].
\label{eq:Hatcav3}
\end{equation}
Without coupling and for zero detuning, the three product states forming the one-photon manifold $\{|u,0\rangle, |e,0\rangle, |g,1\rangle\}$ are degenerate ($|0\rangle$ and $|1\rangle$ denote the photon number states of the cavity). In case of a coupled system, the degeneracy is lifted  and the eigenfrequencies read
\begin{equation}
\omega^0=0 \quad\mbox{and}\quad
\omega^\pm = 
\frac{1}{2}\left(\Delta\pm\sqrt{4 g^2 + \Omega_P^2 +
\Delta^2}\right).
\label{omnnullpm}
\end{equation}
The corresponding eigenstates are
\begin{eqnarray}
|\phi^0\rangle &=& \cos\Theta |u,0\rangle - \sin\Theta 
|g,1\rangle\label{anull1},\\
|\phi^+\rangle &=& \cos\Phi\sin\Theta |u,0\rangle - \sin\Phi |e,0\rangle + 
\cos\Phi\cos\Theta |g,1\rangle,\nonumber\\
|\phi^-\rangle &=& \sin\Phi\sin\Theta |u,0\rangle + \cos\Phi |e,0\rangle + 
\sin\Phi\cos\Theta |g,1\rangle,\nonumber
\end{eqnarray}
where the mixing angles $\Theta$ and $\Phi$ are given by
\begin{equation}
\tan\Theta = \frac{\Omega_P}{2 g} \quad\mbox{and}\quad \tan\Phi = 
\frac{\sqrt{4 g^2 +
\Omega_P^2}}{\sqrt{4 g^2+\Omega_P^2+\Delta^2}-\Delta}.\label{angles}
\end{equation}
Recently, we have proposed \cite{Kuhn99} and demonstrated \cite{Hennrich00,Kuhn01,Kuhn02} that the dark state $|\phi^0\rangle$ can be used to generate photons in the cavity in a deterministic way. To do so, the interaction of a single atom with the cavity must be strong when the interaction starts, so that
\begin{equation}
|\langle u,0|\phi^0\rangle| = 1 
\quad\mbox{for}\quad
2g\gg\Omega_{P}.
\end{equation}
Provided the system's state vector, $|\Psi\rangle$, adiabatically follows the dark state, $|\phi^0\rangle$, throughout the interaction, the contribution of $|g,1\rangle$ increases with rising pump Rabi frequency $\Omega_{P}$, so that the transient photon-emission rate from the cavity reads
\begin{equation}
R_{emit}=2\kappa |\langle g,1|\phi^0\rangle|^2 \mbox{tr}(\rho)=
2\kappa\frac{\Omega_{P}^2}{4g^2+\Omega_{P}^2}\mbox{tr}(\rho),
\end{equation}
where $\kappa$ is the field decay rate of the cavity and $\rho=|\Psi\rangle\langle\Psi|$ the density matrix of the considered three-level system, which is not closed. Hence, $\mbox{tr}(\rho)$ is not one, but denotes the population remaining in the coupled system at any given moment. With the other eigenstates, $|\phi^\pm\rangle$, not being populated, transverse spontaneous emission losses from the excited state $|e\rangle$ do not occur, and the overall photon emission probability is
\begin{equation}
P_{emit}=\int_{0}^\tau dt \: R_{emit} \to 1
\quad\mbox{for}\quad
2\kappa\tau\left\langle\frac{\Omega_{P}^2}{4g^2+\Omega_{P}^2}\right\rangle \gg 1,
\end{equation}
where $\langle\ldots\rangle$ denotes the average of its argument in the relevant time interval, $[0\ldots\tau]$. Obviously, the process can reach a photon-generation efficiency close to unity, provided adiabatic following is assured. Therefore,  the set of parameters $\{\Omega_{0},g_{0},w_{C},w_{P},\Delta,\delta_{X}\}$ must be chosen in such a way that the adiabaticity criterion
\cite{Messiah59} 
\begin{equation}
|\omega^0-\omega^\pm|\gg|\langle\phi^\pm|\frac{d}{dt}|\phi^0\rangle|=|\dot{\Theta}| 
\label{adicrit}
\end{equation}
is met throughout the whole process. This requires a careful adjustment  of the displacement, $\delta_{X}$, between pump laser beam and cavity axis.

\begin{figure}
\includegraphics[width=8.5cm]{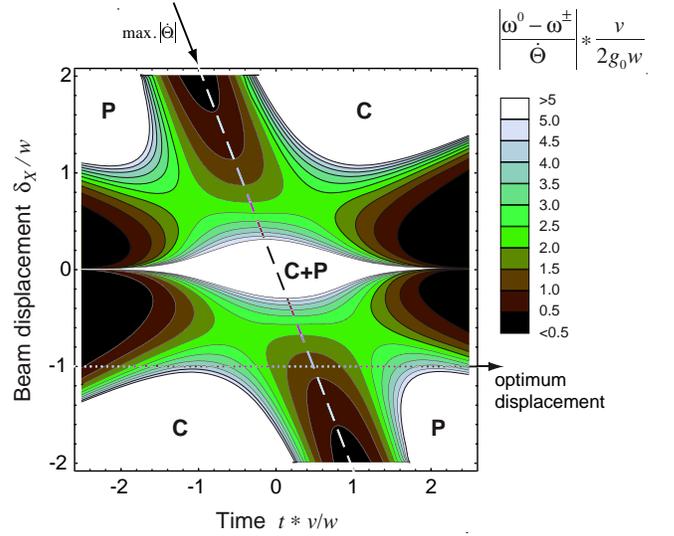}
\caption{\label{adic}
Regimes of high and low adiabaticity: The graymap shows $|(\omega^0-\omega^\pm)/\dot{\Theta}|$ as a function of time and beam displacement for $\Delta=0$, $w_{C}=w_{P}\equiv w$ and $\Omega_{0}=2g_{0}$. In the areas marked with \textbf{C} and/or \textbf{P}, the atom is interacting with the cavity and/or the pump laser beam, respectively. In the dark regions, adiabaticity is not assured. The dashed line indicates the position of the maxima of $|\dot{\Theta}|$, where the probability for non-adiabatic losses is highest.  All atom trajectories, regardless of $\delta_{X}$, cross this line. The dotted horizontal line indicates the optimal atom trajectory that leads to the maximal photon emission probability in our experiment.}
\end{figure}

Figure \ref{adic} shows $|(\omega^0-\omega^\pm)/\dot{\Theta}|$ as a function of $t$ and $\delta_{X}$. The smaller this ratio, the higher the probability to have a non-adiabatic evolution. Obviously, for any choice of $\delta_{X}$, such situations are encountered. The evolution is always non-adiabatic long before and long after the atom interacts with laser beam and/or cavity. However, since the coupling constant and the Rabi frequency are both weak in these regimes and $\Theta$ changes only slowly, the population in the atomic bare states is not affected. A very crucial situation is met when $|\dot{\Theta}|$ reaches its maximum, i.e. when the atom crosses the dashed line in fig.\,\ref{adic}, which lies  between cavity axis and pump beam. For  $|\delta_{X}|>\max(w_{C},w_{P})$, the level splitting $|(\omega^0-\omega^\pm)|$ is too small to assure adiabaticity and losses to other eigenstates occur. Only in the intermediate regime, where the pump beam partially overlaps the cavity mode, i.e. for $|\delta_{X}/2|\leq \min(w_{C},w_{P})$, adiabaticity is assured even for large $|\dot{\Theta}|$, and the behaviour of the system is predetermined by the projection of the initial state $|u,0\rangle$ onto the eigenstates as soon as the adiabatic regime is entered. In this case, three major scenarios must be distinguished:
(a) $\delta_{X}>0$. The atom first interacts with the pump beam, the initial state projects onto $|\phi^\pm\rangle$, and the photon is lost by transverse spontaneous emission.
(b) $\delta_{X}\approx 0$. The dark state $|\phi^0\rangle$ is partially populated, so that the probability of a photon emission from the cavity might reach 50\%.
(c) $\delta_{X}<0$. The atom first interacts with the cavity, the initial state $|u,0\rangle$ projects onto the dark state $|\phi^0\rangle$, so that no losses occur except the desired photon emission from the cavity with a probability approaching 100\%.

\begin{figure}
\includegraphics[width=8.5cm]{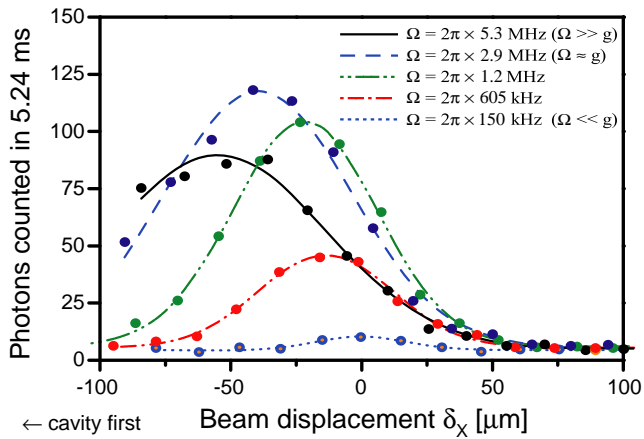}
\caption{\label{kfig2} Number of photons emitted from the cavity in 5.24\,ms as a function of the  displacement $\delta_{X}$ of the pump beam from the cavity axis (along the atom's trajectories) for different pump  intensities. A counter-intuitive interaction sequence (cavity first, pump later) is realized for negative $\delta_{X}$. Note that the indicated Rabi frequencies result from the measured beam intensities and an assumed pump-beam waist of $w_{P}=44\,\mu$m, which gives the best agreement with numerical simulations (see Fig.\,\protect\ref{kfig3}).}
\end{figure}

Figure \ref{kfig2} shows the number of photons emitted from the cavity as a function of pump beam displacement $\delta_X$ for five different pump intensities. The photons are counted during $5.24\,$ms while a single cloud of atoms passes through the cavity. Pump beam and cavity are Raman resonant, with a common detuning of $\Delta=10\,$MHz from the direct atomic transitions. The pump beam displacement is adjusted using a piezoelectric mirror assembly, and beam position and waist are monitored by a CCD camera. The trace that belongs to the smallest pump intensity is used to calibrate the $\delta_X$-origin, assuming that the small Rabi frequency gives rise to an adiabatic evolution only for $\delta_X=0$, so that the peak photon number is found there. From this position,  the  peak emission shifts towards negative $\delta_X$ with increasing Rabi frequency, since the level splitting $|\omega^{0}-\omega^{\pm}|$ and therefore the adiabatic regime both increase with $\Omega_{0}$. This allows to pull pump beam and cavity further apart without loosing adiabaticity, so that the fraction of the initial state that projects onto the dark state $|\phi^0\rangle$ when the adiabatic regime is entered increases as well. The data also reveal that a significant overlap between pump beam and cavity mode is mandatory, otherwise adiabaticity is not assured when the atom reaches the point of maximum $|\dot{\Theta}|$, which is located between cavity mode and pump beam axis, so that the number of emitted photons decreases again for $\delta_{X}\longrightarrow -\infty$. In our experiment, the optimum is found for $\Omega_{P}\approx g_{0}$ and $\delta_{X} \approx -(w_{C}+w_{P})/2$.

\begin{figure}
\includegraphics[width=8.5cm]{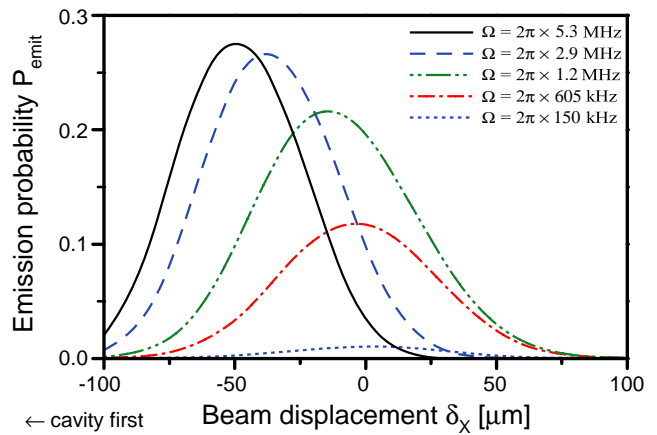}
\caption{\label{kfig3} Simulation of the photon emission probability averaged over all possible points of impact as a function of the  displacement $\delta_{X}$ of the pump beam from the cavity axis for different pump intensities, with $\{g^{max}_{0}, \Delta, \kappa, \gamma\} = 2\pi\{2.5, 10, 1.25, 6\}\,$MHz, cavity waist $w_{C}=35\,\mu$m and pump waist $w_{P}=44\,\mu$m.}
\end{figure}

A comparison of the experimental results with a numerical simulation of the process must take into account that the atomic trajectories are not controlled. Therefore the effective atom-cavity coupling depends on the random `point of impact', $\vec{r}=(y,z)$, of every single atom with respect to the cavity's mode function, so that
\begin{equation} 
g_{0}(y,z)\approx g^{max}_{0} \cos(2\pi z/\lambda) \exp\left(-(y/w_{C})^2\right).
\end{equation}
The variation of $w_{C}$ along $z$ is neglected here, since the Rayleigh length of the cavity mode exceeds the cavity length. The dependence of the photon emission probability $P_{emit}$ on $g_{0}$ is highly non-linear, so that $g_{0}$ cannot be replaced by its mean value prior to a numerical calculation of $P_{emit}$. Instead,we have to average $P_{emit}$ over all possible points of impact, according to
\begin{equation}
\bar{P}_{emit}=\int_{0}^{\lambda}dz\int_{-S_{y}/2}^{S_{y}/2}dy\;
\frac{P_{emit}(g_{0}(y,z))}{\lambda\,S_{y}}.
\label{Pint}
\end{equation}
Note that $S_{y}=100\,\mu$m is the width of a slit aperture that is installed above the cavity. This aperture constrains the atomic trajectories perpendicular to the cavity axis to the interval $[-S_{y}/2\ldots+S_{y}/2]$. To calculate $P_{emit}(g_{0}(y,z))$, we numerically solve the master equation of the coupled system \cite{Kuhn99} and integrate (\ref{Pint}). This calculation takes the cavity-field decay rate, $\kappa$, and the decay rate of the excited atomic level, $\gamma$, into account. 

Figure \ref{kfig3} shows a simulation of the average $\bar{P}_{emit}$ as a function of $\delta_{X}$ for the range of pump intensities investigated in the experiment (see Fig.\,\ref{kfig2}). For the smallest Rabi frequency the maximum photon emission probability is found at $\delta_{X}=0$, which justifies our $\delta_{X}$ calibration. Moreover, the simulation shows the same trend that is observed in the experiment, with the peak emission probability shifting towards negative $\delta_{X}$ with increasing Rabi frequency. However, simulation and experiment do not agree perfectly, and also the experimentally observed photon-number reduction at high pump intensities is not reproduced. These discrepancies lead to the conclusion that either the pump beam deviates from an ideal Gaussian beam in its wings, or that weak stray light of the pump beam (e.g. from the vacuum viewports) also hits the atoms. Both effects give rise to an electronic excitation of the atoms if the pump beam is very intense, which results in an early loss of photons by spontaneous emission into transverse modes.

From the theoretical considerations, the experimental results and the numerical calculations we draw the conclusion that vacuum-stimulated Raman transitions are most effectively driven by an adiabatic passage that results from a counter-intuitive interaction sequence, where the atoms are first coupled to the vacuum field of an empty cavity, stimulating the transition, and then exposed to a pump laser beam. To assure adiabaticity and to avoid losses to other states, a significant overlap of cavity mode and pump beam is required. Optimum conditions are found experimentally for a pump Rabi frequency that equals the maximum atom-cavity coupling constant, $g^{max}_{0}$, and for a beam displacement that equals the average waist of cavity and pump beam.

\begin{acknowledgements}
This work was supported by the focused research program `Quantum Information Processing' of the Deutsche Forschungsgemeinschaft, and by the European Union through the IST (QUBITS) and IHP (QUEST) programs.
\end{acknowledgements}

\end{document}